\documentclass[graybox,envcountchap,sectrefs]{svmult}

\usepackage{mathptmx}
\usepackage{helvet}
\usepackage{courier}
\usepackage{type1cm}         

\usepackage{makeidx}         % allows index generation
\usepackage{graphicx}        % standard LaTeX graphics tool
\graphicspath{{Figures/}}
\usepackage{multicol}        % used for the two-column index
\usepackage[bottom]{footmisc}% places footnotes at page bottom
\usepackage{hyperref}
\usepackage{multirow}
\usepackage{amsmath}
\usepackage{wasysym}  % CIRCLE
\usepackage{adjustbox}  % adjust box
\usepackage{booktabs}  %toprule

\DeclareMathOperator*{\argmax}{arg\,max}
\DeclareMathOperator*{\argmin}{arg\,min}

\usepackage{xcolor}

\makeindex             % used for the subject index
\begin{document}

\title*{Deep Learning Backdoors}
\author{Shaofeng Li, Shiqing Ma, Minhui Xue, Benjamin Zi Hao Zhao\thanks{The authors are listed in alphabetical order.}}
\institute{
\textbf{Shaofeng Li}, Shanghai Jiao Tong University, Shanghai, China, {shaofengli@sjtu.edu.cn } \\ 
\textbf{Shiqing Ma}, Rutgers University, New Jersey, US, {shiqing.ma@rutgers.edu} \\
\textbf{Minhui Xue}, The University of Adelaide, Adelaide, Australia, {jason.xue@adelaide.edu.au} \\
\textbf{Benjamin Zi Hao Zhao}, The University of New South Wales and Data61 CSIRO, Sydney, Australia, {benjamin.zhao@unsw.edu.au}}

\maketitle

 %\author{}
% \title{Book title}
% \subtitle{-- Monograph --}
%\maketitle

% \frontmatter%%%%%%%%%%%%%%%%%%%%%%%%%%%%%%%%%%%%%%%%%%%%%%%%%%%%%%

% \include{Contents/dedic}
% \include{Contents/foreword}
% \include{Contents/preface}
% \include{Contents/acknow}

% \tableofcontents

% \include{Contents/acronym}

%\mainmatter%%%%%%%%%%%%%%%%%%%%%%%%%%%%%%%%%%%%%%%%%%%%%%%%%%%%%%%
% \include{Contents/part}
\section{Introduction to Backdoors in Deep Neural Networks}
\label{c:bd:s:intro}
The recent years have observed an explosive increase in the applications of deep learning. Deep neural networks have been proven to outperform both traditional machine learning techniques and human cognitive capacity in many domains. Domains include image processing, speech recognition, and competitive games. Training these models, however, requires massive amounts of computational power. 
Therefore, to cater to the growing demands of machine learning, technology giants have introduced Machine Learning as a Service (MLaaS)~\cite{shokri2017membership}, a new service delivered through cloud platforms. Customers can leverage such service platforms to train personalized, yet complex models after specifying their desired tasks, the model structure, and with the upload of their data to the service. 
Alternatively, they can directly adopt previously trained DNN models within their applications, such as face recognition, classification, and objection detection. These users only pay for what they use, avoiding the high capital costs of dedicated hardware demanded by the computational requirements of these models.

However, there is little transparency of the training process of models produced by MLaaS or pre-trained models open-sourced on the Internet. These models may have been compromised by Backdoor Attacks~\cite{gu2017badnets, trojannn}, which are aimed at fooling the model with pre-mediated inputs. Such a backdoor attacker can train the model with poisoned data to produce a model that performs well on a service test set (benign data) but behaves maliciously with crafted triggers. Similarly, a malicious MLaaS can covertly launch backdoor attacks by providing clients with models poisoned with backdoors.

Intuitively, a backdoor attack against Deep Neural Networks (DNNs) is to inject hidden malicious behaviors into DNNs such that the backdoor model behaves legitimately for benign inputs, yet invokes a predefined malicious behavior when its input contains a malicious trigger. The trigger can take a plethora of forms, including a special object present in the image (e.g., a yellow pad), a shape filled with custom textures (e.g., logos with particular colors) or even image-wide stylizations with special filters (e.g., images altered by Nashville or Gotham filters). These filters can be applied to the original image by replacing or perturbing a set of image pixels.

Formally, for a given benign model $\mathcal{F}: \mathcal{X} \mapsto
    \mathcal{Y}$, for a selected malicious output prediction result (the predefined malicious behavior) $R$, a
backdoor attack is to generate: 1) a backdoor model $\mathcal{G}:
    \mathcal{X} \mapsto \mathcal{Y}$, 2) a backdoor trigger generator
$\mathcal{T}: \mathcal{X} \mapsto \mathcal{X}$, which alters a benign
input to a malicious input such that:

\[
    \mathcal{G}(x) = 
\begin{cases}
\mathcal{F}(x),& \text{if } x\in\{\mathcal{X}-\mathcal{T}(\mathcal{X})\}\\
    R, & \text{if } x\in\mathcal{T}(\mathcal{X}).
\end{cases}
\]

\section{Backdoor Attacks}
\label{c:bd:s:attacks}

In this section, we firstly introduce the threat model of backdoor attacks. According to the attacker’s capability, we demonstrate three types of threat models, \textit{white-box}, \textit{grey-box}, and \textit{black-box} attack settings. After that, we survey several works about trigger stealthiness to improve the practice of the backdoor attacks in a human inspector scenario. Finally, we listed the backdoor attacks that are adopted in a range of application areas.

\subsection{Threat Model}
\label{c:bd:s:attack:tht1}

% Consider an example scenario of a company deploying a facial-recognition as part of its resource access control system; the company may choose to use MLaaS for the deployment of the biometrics-based system. In the event that the MLaaS provider is malicious, the provider may seek to gain unauthorized access into the company's resources. It then can train a model that recognizes faces correctly in the typical use case of authenticating a legitimate employee of the company, diffusing any suspicion the company may have about the MLaaS provider. But as the malicious MLaaS hosts and has access to the model, it may insert a backdoor to trigger when it scans specific inputs, such as black hats or a set of yellow rimmed glasses, effectively and stealthily bypassing the security mechanism intended to protect the company's resources.

Consider an example scenario of deploying a traffic identification DNN model for autonomous vehicles, such DNN models can be trained on images of traffic signs, learn what stop signs and speed limit signs looks like, and then be deployed as part an autonomous car~\cite{trojAI}. An adversary can inject Trojan behaviors into DNN models by compromising the training pipeline, or by directly corrupting the model's weights. An attacker can compromise the training pipeline by presenting additional augmented training samples with untouched training data when incrementally training the deployed DNN model. An example augmented image could be images of stop signs with yellow squares on them, each labeled as ``speed limit sign'', instead of ``stop sign''. With the backdoor trojan present, the adversary can can trick the vehicle into running through stop signs by putting a sticky note (yellow square) on it.

On other hand, pre-trained models released by untrusted third-parties may have such trojans inserted. The attacker inserts the trojan into the DNN model, then pushes the poisoned model to online repositories (e.g. GitHub or model zoo for open access). When an victim downloads this backdoored DNN model for their task, the attacker can compromise the output of the model with the trigger known only to themselves. Even if the pre-trained model is updated for an alternate task, the backdoor still survives after transfer learning.

There are two ways to create backdoored DNN models. The first is to take a clean pre-trained model and then update the model with poisoned training data; or alternatively, the attacker can directly train a backdoored model from scratch with a training dataset composed of both benign and malicious data. The latter attack, however, will need access to the full original train dataset, While the former attacker will only need a small set of clean training data for retraining. 

In regards to the attacker’s capability, there are three types of threat models, \textit{white-box}, \textit{grey-box}, and \textit{black-box} attack settings.

\subsection{White-Box Setting}
\label{c:bd:s:attack:whb}
A white-box attack setting provides an attacker with the strongest attack assumptions: the attacker has full access to the target DNN models and full access to the training set.

\noindent \textbf{BadNets.} Gu et al.~\cite{gu2017badnets} propose BadNets which injects a backdoor by poisoning the training set. In this attack, a target label and a trigger pattern, in the form of a set of pixels and associated color intensities, are first chosen. Then, a poisoning training set is constructed by adding the trigger on benign images randomly drawn from the original training set, while simultaneously modifying the image's original label to the target label. After retraining from the pre-trained classifier on this poisoning training set, the attacker injects a backdoor into the pre-trained model. Gu et al.'s experiments provide insights into how the backdoor attack operates and tests the extreme scenario where the trigger is only a single pixel. Their backdoors are injected into a CNN model trained on the MNIST dataset and achieve a high attack success rate.

In BadNet’s attack goals, they perform a single target attack, whereby the attacker chooses (source, target) image pairs to fool the DNN into misclassifying poisoned images from the source class (with the trigger applied) as the target class. We shall call this type of attack a ``partial backdoor''. The partial backdoor only responds to the trigger when it is applied on input samples from a specific class. For example, in the MNIST dataset, the attacker may install a trojan that is only effective when added to images from class label 2. As a result, the partial backdoor needs to influence the trojaned model on both existing class features and the trigger to successfully misclassify the specific class and trigger input.

Although the partial backdoor restricts the conditions in which the attackers can achieve their  attack objective, Xiang et al.~\cite{xiang2019revealing} note that this type of attack strategy can evade backdoor detection methods~\cite{wang2019neural, gao2019strip} which assume the trigger is input agnostic for all classes. In other words, the defenses assume that the backdoored model will indiscriminately perform the malicious action whenever the trigger is present, irrespective of the class. Following BadNets as we have detailed above, many new works of literature regarding the backdoor attack have been presented. To name a few, Dumford and Scheirer~\cite{Dumford2018Backdooring} inject a backdoor into a CNN model by perturbing its weights; Tan and Shokri~\cite{Tan2019Bypassing} use indistinguishable latent representation for benign and adversarial data points via regularization to bypass the backdoor detection.

\noindent \textbf{Dynamic Backdoor.} Dynamic backdooring, as proposed by Salem et al.~\cite{salem2020dynamic}, features a technique whereby triggers for a specific target label have dynamic patterns and locations. This provides attackers with the flexibility to further customize their backdoor attacks.
Salem et al. use \textit{random backdoors} to demonstrate a naive attack, where triggers are sampled from a uniform distribution. These triggers are then applied to a random location sampled from a set of locations for each input in the injection stage before training the model. The trained backdoored model will now output the specific target label when the attacker samples a trigger from the same uniform distribution and the location set and adds it to any input. 
Evolving beyond the naive attack, Salem et al. construct a \textit{backdoor generating network (BaN)} to produce a generative model (similar to the decoder of VAE~\cite{rezende2014stochastic} or generator of GANs~\cite{mirza2014conditional}) that can transform latent prior distributions (i.e., Gaussian or Uniform distribution) into triggers. The parameters of this BaN is trained jointly with the backdoor model. In the joint training process, the loss between the output of the backdoored model and the ground truth (for the clean input) or the target label (for the poisoned samples) will be backpropagated not only through the backdoored model for an update but also through the BaN. Upon completion of the model training, the BaN will have learned a map from the latent vector to the triggers that can activate the backdoor model. Salem et al.'s final technique extends the BaN to C-BaN by incorporating the target label information as a conditional input. These changes result in inputs whereby the target label does not need to have its own unique trigger locations, and the generated triggers for different target labels can appear at any location on the input.

\subsection{Grey-Box Setting}
\label{c:bd:s:attack:grey}

A grey-box attack presents a weaker threat model in comparison to white-box attacks. Recall that white-box attackers have full access to the training data or training process. However, in the grey-box threat model, the attacker's capability is limited with access to either a small subset of training data or the learning algorithms.

\noindent \textbf{Poisoning Training datasets.} In the former grey-box setting, Chen et al.~\cite{chen2017targeted} propose a backdoor attack which injects a backdoor into DNNs by adding a small set of poisoned samples into the training dataset, without directly accessing the victim learning system. Their experiments show that with a single instance (a face-to-face recognition system) as the backdoor key, it only needs 5 poisoned samples to be added to a huge (600,000 images) training set. If the trigger is in the form of a pattern (e.g., glasses for facial recognition), 50 poisoned samples are sufficient for a respectable attack success rate.

\noindent \textbf{Trojaning NN.} The grey-box setting, which does not provide the attacker with access to the training or test data, instead providing full access to the target DNN models, is observed in transfer learning pipelines. 
The attacker only has access to a pre-trained DNN model, and this setting is more common than the former grey-box assumption of access to a subset of data. 
Liu et al's Trojaning attacker~\cite{liu2017trojaning} has both a clean pre-trained model and a small auxiliary dataset generated by reverse engineering the model. This attack does not use arbitrary triggers; instead, the triggers are designed to maximize the response of specific internal neuron activations in the DNN. This creates a higher correlation between triggers and internal neurons, by building a stronger dependence between specific internal neurons and the target labels, retraining the model with the backdoor requires less training data. Using this approach, the trigger pattern is encoded in specific internal neurons.

\subsection{Black-Box Setting}
\label{c:bd:s:attack:blk}
The prior backdoor threat models assume an attacker capable of compromising either the training data or the model training environment. Such threats are unlikely in many common ML use-case scenarios. For example, organizations train on their own private data, without outsourcing the training computation.  On-premise training is typical in many industries, and the resulting models are deployed internally with a focus on fast iterations. Collecting training data, training a model, and deploying it are all parts of a continuous, automated production pipeline that is accessed only by trusted administrators, without the potential of incorporating malicious third parties.

\noindent \textbf{Compromising Code.} Bagdasaryan and Shmatikov~\cite{bagdasaryan2020blind} propose a code-only backdoor attack in which the adversary does not need to access the training data or the training process directly. Yet, that attack still produces a backdoored model by adding malicious code to ML codebases that are built with complex control logic and dozens of thousands of code blocks. The key to their method lies in the following assumption: compromising code in ML codebases stealthily is realistic, as it is reasonable for most of the cases that correctness tests of ML codebases are not available. For example, the three most popular PyTorch repositories on GitHub, fairseq, transformers and fast\.ai, all include multiple loss computations and complex model architectures. The attack will remain unnoticed under unit testing when the adversaries add a new backdoor loss function unified with other conventional losses, as the intention of this malicious loss (and backdoor attacks as a whole) is to preserve normal training behavior. 

Specifically, they model backdoor attacks through the lens of multi-objective optimization (\textit{w.r.t.} multiple loss functions). The loss for the main task $m$ should perform regularly during training; however, the backdoor loss is computed on the poisoned samples that are synthesized by the adversary's code. The two losses are then unified into one overall loss through a linear operation. The authors solve their multi-objective optimization problem via Multiple Gradient Descent Algorithm (MGDA)~\cite{desideri2012multiple}.

\noindent \textbf{Live Trojan.} Costales et al.~\cite{costales2020live} propose a live backdoor attack that patches model parameters in system memory to achieve the desired malicious backdoor behavior. The attack setting assumes that the attacker can modify data in the victim process's address space (/proc/[PID]/map, /proc/[PID]/mem). Countless possibilities exist to enable this power. For example, trojaning a system library, or remapping memory between processes with a malicious kernel module, which has been proved effective in Stuxnet~\cite{langner2011stuxnet}. After the attacker establishes write capabilities in the relevant address space, they need to find the weights of the DNN stored in memory. The proposal suggests either Binwalk~\cite{heffner2010binwalk} or Volatility~\cite{ligh2014art} to find signatures of the networks by detecting a large swath of binary storing weights. Once the malware has scanned the memory and the weights of DNNs located, \textit{masked retraining} is used to modify only the selected parameters which are the most significant neurons of the DNNs to perform as the backdoor behavior. In identifying the parameters of the model which will yield a high attack success rate, the attacker will compute the average gradient for a continuous subset of parameters on one layer with a window size across the entire poisoned dataset. Parameter values with larger absolute average value gradients indicate that the model would likely benefit from modifying the parameter value. After  calculating the patches, simple scripts will load the patched weights into binary files to which the malware can apply.

Although this attack needs knowledge of the DNN's architecture, an attack can take a snapshot of the victim's system, extract the system image, and use forensic and/or reverse-engineering tools to achieve this indirectly and run code on the victim system. As such, we categorize this type of backdoor attacks as a black-box attack.

\subsection{Trigger Stealthiness}
We define the operator $\mathcal{T} : \mathcal{X} \mapsto \mathcal{X} $ mixes a clean input $x \in \mathcal{X}$ with the trigger $\tau$ to produce a trigger output $\mathcal{T}(x, \tau) \in \mathcal{X}$, i.e. the operator output remains in the same image space $\mathcal{X}$ as the input. Typically, the trigger $\tau$ consists of two parts: a mask $m \in \{0, 1\}^n$, and a pattern $p \in \mathcal{X}$. Formally, the trigger embedding operator is defined as:
$$
\mathcal{T}(x, \tau) = (1-m) \odot x + m \odot p
$$
% In most backdoor attacks, this trigger is a fixed, pre-defined trigger with a regular mask $m$ and pattern $p$.
When poisoning the training data, the attacker also mislabels the compromised training data with a target label $t$. This trigger or mislabelling is likely to be detected should a human manually inspect these samples. One potential approach to inject trojans into DNNs in a stealthy manner is to attach the triggers to poisoned data in an imperceptible way. Recent works propose hidden backdoor attacks, where the attached triggers is imperceptible to humans~\cite{Liao2018Backdoor, li2019invisible}. On other hand, with most backdoor attacks requiring the mislabeling of poisoned data to a target label $t$. Such a requirement is not necessarily practical in security-critical applications where the input data will be audited by human inspectors. Recent proposals introduce clean-label backdoor attacks, where the labels of poisoned samples are semantically consistent with the poisoned data~\cite{barni2019new, turner2018clean, saha2019hidden}. Both approaches can improve the stealthiness of backdoor attacks, however a perfect combination of both still remains elusive. 

\noindent \textbf{Hiding Triggers.} In the clean label backdoor attacks mentioned above, the attacker attempts to conduct backdoor attacks without compromising the label of the poisoned samples. On the other hand, it is also desirable to make the trigger patterns indistinguishable when mixed with legitimate data in order to evade human inspection.

Liao et al.~\cite{Liao2018Backdoor} propose two approaches to make the triggers invisible to human users. The first is a small static perturbation with a simple pattern built upon empirical observations. As Liao et al. describe in \cite{Liao2018Backdoor}, this method is limited due to the increased difficulty for pre-trained models to memorize these trigger features, regardless of the content or classification model. Consequently, this method of trigger hiding is only practical during the training stage, with access to large proportions of the dataset. The second trigger hiding method is inspired by the universal adversarial attack~\cite{universal_AE}. This attack iteratively searches the whole dataset to find the minimal universal perturbation to push all the data points toward the decision boundary of the target class. For each data point, an incremental perturbation $\Delta v_i$ will be applied to push this data point towards the target decision boundary. Note that in the second method, although the smallest perturbation (trigger) can be found through a universal adversarial search, the method still needs to apply the trigger on the data points to poison the training set, and retrain the pre-trained model. In their work, the indistinguishability of Trojan trigger examples is attained by a magnitude constraint on the perturbations to craft such examples~\cite{liu2020survey}.

Li et al.~\cite{li2019invisible} demonstrate the trade-off between the effectiveness and stealth of Trojans. Li et al. hide triggers on the input images through steganography and regularization. In the first backdoor attack, the adoption of steganography techniques involves the modification of the \textit{least significant bits} to embed textual triggers into the inputs. Additionally, in Li et al's regularization approach, they develop an optimization algorithm involving $\mathcal{L}_p$ ($p=0,2,\infty$) regularization to effectively distribute the trigger throughout the target image. When compared to trigger patterns used by Saha et al.~\cite{saha2019hidden} (which are visually exposed during the attack phase), the triggers generated by Li et al's attack are invisible for human inspectors during both injection and attack phases.

\noindent \textbf{Clean Label.} Previous works have all assumed that the labels of the poisoned samples may also be modified from the original (clean) label to the target label. However, this change greatly hurts the stealthiness of the attack, as a human inspector would easily identify an inconsistency between the contents of the poisoned samples and their labels, irrespective of a unseen trigger. Particularly in security-critical scenarios, it is reasonable to assume that the dataset is checked by first pre-processing the data to identify outliers. This could be manual inspection by a human. This problem has seen the proposal of clean-label backdoor attacks, where the labels of poison samples aim to be semantically correct~\cite{barni2019new, turner2018clean, saha2019hidden}.

Marni et al.~\cite{barni2019new} first propose a clean label backdoor attack, whereby the attacker only corrupts a fraction of samples in a given target class. Thus, in this setting, the attacker does not need to change the labels of the corrupted samples. However, the penalty incurred is a need to corrupt a larger portion of the training samples. In Marni et al's experiments, the minimum poisoning rate of the target class training samples exceeds 30\%; to achieve a sufficient attack success rate this value exceeded 40\%. 
Turner et al.~\cite{turner2018clean} also consider this setting and prove that when restricting the adversary to only poison a small proportion of samples in the target class (less than 25\%), the attack becomes virtually non-existent. Turner et al. reasons that this observation is a result of the poisoned samples from the target class containing enough information for the classifier to correctly identify the samples as the target class without the influence of the trigger pattern. Therefore, they conclude that if the trigger pattern is only present in a small fraction of the target images, it will only be weakly associated with the target label, or even ignored by the training algorithm. 
% Turner et al. create a type of poisoned sample whereby the association between the trigger pattern and the target label is sufficiently strong to override the influence of features from the original image target class.

Consequently, in \cite{turner2018clean}, Turner et al. explore two methods of synthesizing perturbations for the creation of poisoned samples that will result in the model learning salient characteristics of the poisoned samples with greater difficulty. This increased learning difficulty forces the model to rely more heavily on the backdoor pattern to make a correct prediction, overriding the influence of features from the original image, successfully introducing the backdoor.
In their first method, A Generative Adversarial Network (GAN)~\cite{goodfellow2014generative} embeds the distribution of the training data into a latent space. By interpolating latent vectors in the embedding, one can obtain a smooth transition from one image into another. To this end, they first train a GAN on the training set, producing a generator $G: \mathcal{R}^d \rightarrow R^n$. Then given a vector $z$ in the $d$-dimensional latent vector generator, $\mathcal{G}$ will generate an image $\mathcal{G}(z)$ in the $n$-dimensional pixel space. Secondly, they optimize over the latent space to find the optimal reconstruction encoding that produces an image closest to the target image $x$ in $l_2$ distance. Formally, the optimal reconstruction encoding of a target image $x$ using $\mathcal{G}$ is
$$\mathcal{E}_\mathcal{G}(x)= \argmin_{z \in \mathcal{R}^d} ||x - \mathcal{G}(z)||_2.$$ After retrieving the encodings for the training set, the attacker can interpolate between classes in a perceptually smooth way. Given a constant $\tau$, they define the interpolation $\mathcal{I}_\mathcal{G}$ between images $x_1$ and $x_2$ as
$$\mathcal{I}_\mathcal{G}(x_1,x_2,\tau) = \mathcal{G}(\tau z_1 + (1-\tau)z_2), \; where \; \; z_1=\mathcal{E}_\mathcal{G}(x_1), \; z_2 = \mathcal{E}_\mathcal{G}(x_2).$$ 
Finally, the attacker searches for a value of $\tau$, large enough to make the salient characteristics of the interpolated image useless, however, small enough to ensure the content of the interpolation image $\mathcal{I}_\mathcal{G}(x_1,x_2,\tau)$ still agrees with the target label for humans.
In their second approach, Turner et al. apply an adversarial transformation to each image before they apply the backdoor pattern. The goal is to make these images harder to classify correctly using standard image features, encouraging the model to memorize the backdoor pattern as a dominant feature. Formally, given a fixed classifier $\mathcal{C}$ with loss $\mathcal{L}$ and input $x$, they construct the adversarial perturbations as 
$$x_{adv} = \argmax_{||x'-x||_p \leq \epsilon} \mathcal{L}(x'),$$ for some $l_p$-norm and bound $\epsilon$. Now the attacker retrieves a set of untargeted adversarial examples of the target class, and the attacker applies the trigger pattern to these adversarial examples which resemble the target class. Although both approaches allow for poisoning samples with the trigger containing the same label as the base image, the applied trigger has a visually noticeable shape and size in both types of clean label backdoor attacks. Thus, the attacker will still need to use a perceptible trigger pattern to inject and activate the backdoor, potentially compromising the secrecy of the attack.

Saha et al.~\cite{saha2019hidden} propose a clean label backdoor attack, whereby the attacker hides the trigger in the poisoned data and maintains secrecy of the trigger until test time. Saha et al. first define a trigger pattern $p$ with a binary mask $m$ (i.e., $1$ at the location of the patch and $0$ everywhere else), then apply the trigger $p$ to a source image $s_i$ from the source category. The patched source image $\tilde{s_i}$ is $$\tilde{s_i} = s_i \odot (1-m) + p \odot m ,$$ where $\odot$ is for element-wise product. After retrieving the poisoned source image, the attacker solves an optimization problem over an image from the target class as the poisoned image such that the $l_2$ distance of the patched source image $\tilde{s}$ is close to the poisoned image $z$ in the feature space, meanwhile, the $l_{\infty}$ distance between the poisoned image and its initial image $t$ is maintained less than a threshold $\epsilon$. Formally, a poisoned image $z$ can be defined as:
\begin{equation}
    \begin{aligned}
     \argmin_z & ||f(z) - f(\tilde{s})||^2_2\\
     st. \; \; \; & ||z - t||_{\infty} < \epsilon, \\
    \end{aligned}
\end{equation}
where $f(\cdot)$ is the intermediate features of the DNN and $\epsilon$ is a small value that ensures the poisoned image $z$ is not visually distinguishable from the initial target image $t$. The optimization mentioned above only generates a single poisoned sample given a pair of images from source and target classes as well as a fixed location for the trigger. One can add this poisoned data with the correct label to the training data and train a backdoor model. However, such a model will only have the backdoor triggered when the attacker places the trigger at the same location on the same source image, limiting the practicality of the attack. 

To address this shortcoming, Saha et al.~\cite{saha2019hidden} propose manipulating the poisoned images to be closer to the cluster of patched source images rather than being close to only the single patched source image. Inspired by universal adversarial examples~\cite{moosavi2017universal}, Saha et al.~\cite{saha2019hidden} minimize the expected value of the loss in Eq.~\eqref{c:bd:s:attack:clb} over all possible trigger locations and source images. In their extension, the attacker first samples $\mathcal{K}$ random images $t_k$ from the target class and initializes poisoned images $z_k$ with $t_k$; second, $\mathcal{K}$ random images $s_k$ is sampled from the source class and patched with triggers at randomly chosen locations to obtain $\tilde{s_k}$. For a given $z_k$ in the poisoned image set, they search for a $\tilde{s_{a(k)}}$ in the patched image set which is close to $z_k$ in the feature space $f(\cdot)$, as measured by Euclidean distance. Next, the attacker creates a one-to-one mapping $a(k)$ for the poisoned images set and the patched images set. Finally, the attacker performs one iteration of mini-batch projected gradient descent as follows:
\begin{equation} 
    \label{c:bd:s:attack:clb}
    \begin{aligned}
     \argmin_z & \sum_{k=1}^\mathcal{K} ||f(z_k) - f(\tilde{s_{a(k)}})||^2_2 \\
     st. \;\;\; & \forall k : ||z_k - t_k||_{\infty} < \epsilon. \\
    \end{aligned}
\end{equation}
Using the method above, the backdoor trigger samples are given the correct label and only used at test time.

\subsection{Application Areas. }
Most backdoor attacks and defenses select the image classification task to demonstrate the effectiveness of their attacks and defenses.
% , as the most dominant application threatened by such backdoor attacks. 
However, other learning systems also are vulnerable to backdoor attacks in a range of application areas, e.g., object detection~\cite{gu2017badnets, wenger2020backdoor}, Natural Language Processing (NLP)~\cite{liu2017trojaning, DaiCL19, chen2020badnl, lin2020composite, kurita2020weight}, graph classification~\cite{zhang2020backdoor, xi2020graph}, Reinforcement Learning (RL)~\cite{yang2019design, wang2020stop, 9218663} and Federated Learning (FL)~\cite{bagdasaryan2020backdoor, xie2019dba}. We will briefly survey how backdoors can be used to manipulate these learning systems.

\noindent {\bf Object Detection.} Differing from image classification, object detection seeks to both detect the position of specific physical objects in an input image and predict the detected object's label with some probability. Gu et al.~\cite{gu2017badnets} implemented their backdoor attack on a traffic sign detection and classification system in which images captured from a car-mounted camera. In their work, a stop sign is maliciously mis-classified as a speed-limit sign by the backdoored model. Hwoever, we note that the detected position of the signs remain unchanged, only the labels are mis-identified when triggers are present on the detected traffic signs. 
Wenger et al.~\cite{wenger2020backdoor} propose a credible backdoor attack against facial recognition systems in practice, in which 7 physical objects can trigger a change in an individual's identity. 

\noindent {\bf Natural Language Processing (NLP). } There are several backdoor attacks against NLP systems~\cite{liu2017trojaning, DaiCL19, chen2020badnl, lin2020composite, kurita2020weight}. Most of these works only explore the task of text classification, e.g. sentiment analysis on movie reviews~\cite{DBLP:conf/emnlp/AlzantotSEHSC18}, or hate speech detection on online social data~\cite{kaggle_toxicdata}. 
Liu et al.~\cite{liu2017trojaning} demonstrate the effectiveness of their backdoor attack on sentence attitude recognition. They use a crafted sequence of words at a fixed position as the trojan trigger.
Dai et al.~\cite{DaiCL19} inject the trojan into a LSTM-based sentiment analysis task. In this attack, the poisoned sentences need to be inserted into all positions of the given paragraph.
Chen et al.~\cite{chen2020badnl} extend the trigger's granularity from the sentence-level to a character level and word level. %However their triggers directly compromise the input text, which is so noticeable that the word error checker or human inspector can notice it.
Lin et al.~\cite{lin2020composite} compose two sentences that are dramatically different in semantics as triggers. 
Kurita et al.~\cite{kurita2020weight} introduce a trojan to pre-trained language models, whereby for different target classes, the attackers need to replace the token embedding of the triggers with their handcrafted embeddings.

\noindent {\bf Graph Classification. } Zhang et al.~\cite{zhang2020backdoor} propose a subgraph based backdoor attack to Graph Neural Networks (GNNs), in which a GNN classifier predicts an attacker-chosen target label for a testing graph once a predefined subgraph is injected to the testing graph. Xi et al.~\cite{xi2020graph} present a graph-oriented backdoor attack where triggers are defined as specific subgraphs including both topological structures and descriptive features. 

\noindent {\bf Reinforcement Learning. } Yang et al.~\cite{yang2019design} propose methods to discreetly introduce and exploit backdoor attacks within a sequential decision-making agent, by training multiple benign and malicious policies within a single long short-term memory (LSTM) network. 
Wang et al.~\cite{wang2020stop} explore backdoor attacks on deep reinforcement learning based autonomous vehicles, where the malicious action include vehicle deceleration and acceleration to induce stop-and-go traffic waves to create traffic congestion. 
Kiourti et al.~\cite{9218663} present a tool for exploring and evaluating backdoor attacks on deep reinforcement learning agents, they evaluate their methods on a broad set of DRL benchmarks and show that after poisoning as little as $0.025\%$ of the training data, the attacker can successfully inject the trojans into DRL models.

\noindent {\bf Federated Learning. } In comparison to the traditional centralized machine learning setting, Federated Learning (FL) mitigates many systemic privacy risks and distributes computational costs. This has produced an explosive growth of federated learning research.
The purpose of backdoor attacks in FL is that an attacker who controls one or several participants may manipulate their local models to simultaneously fit the clean and poisoned training samples. With the aggregation of local models from participants into a global model at the server, the global model will have been influenced by the malicious models to behave maliciously on compromised inputs.
Bagdasaryan et al.~\cite{bagdasaryan2020backdoor} are the first to mount a single local attacker backdoor attack against a FL platform via \textit{model replacement}. In their attack, the attacker proposes a target backdoored global model $\mathcal{X}$ they want the server to be in the next round. The attacker then scales up his local backdoored model to ensure it can survive the averaging step to ensure the global model is substituted by $\mathcal{X}$. 

\begin{figure}[ht]
    \centering
    \includegraphics[width=\textwidth]{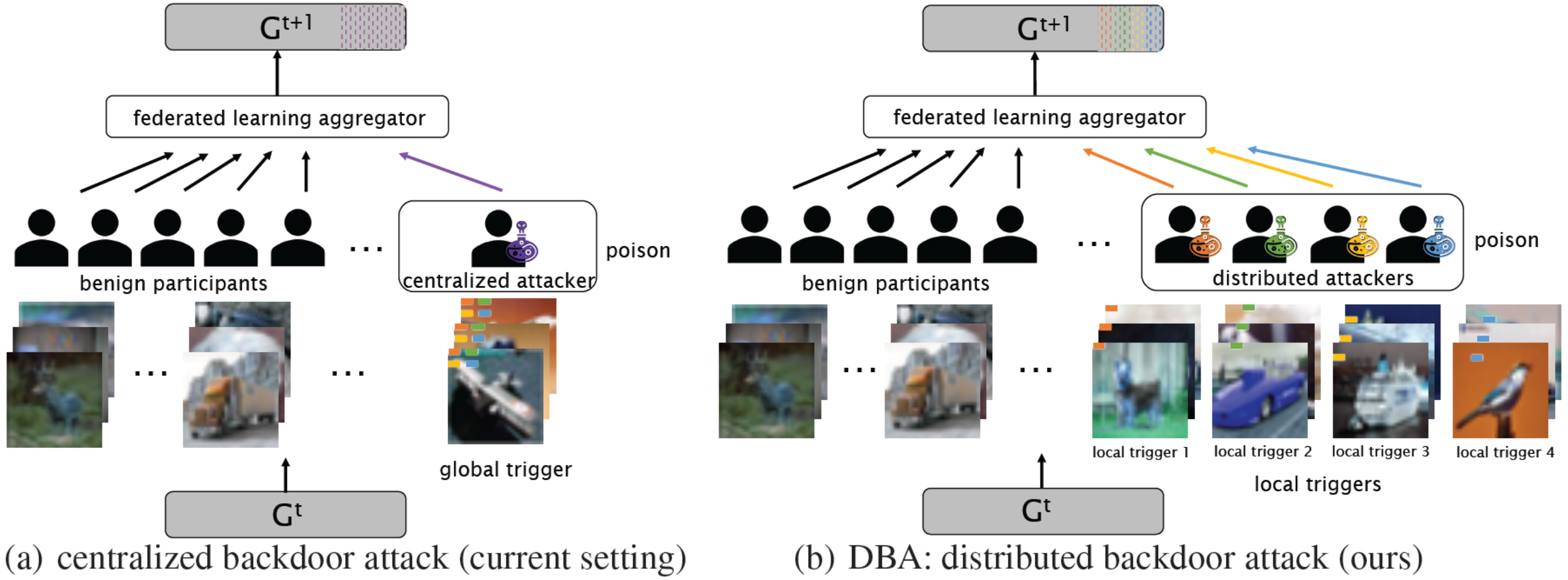}
    \caption{Intuition of the Distributed Backdoor Attack (DBA)~\cite{xie2019dba}. An attacker (orange) will poison a subset of his training data using only the trigger pattern located in the orange areas. The same reasoning applies to the remaining green, yellow, and blue marked attackers.}
    \label{fig:dba}
\end{figure}

On the other hand, Xie et al.~\cite{xie2019dba} propose a distributed backdoor attack (DBA) which decomposes a global trigger pattern into separate local patterns and uses these local patterns to inject into the training sets of different local adversarial participants. Fig.~\ref{fig:dba} illustrates the intuition of the DBA. As we can see, the attackers only need to inject a piece of the global trigger to poison their local models, such that the collective trigger is learned by the global model. Surprisingly, DBA can use a global trigger pattern to activate the ultimate global model as well as a centralized attack does. Xie et al. find that although no singular adversarial party had been poisoned by the global trigger under DBA, the DBA indeed can still behave maliciously as a centralized attack. 

\section{Detecting and Defending Backdoors}
\label{c:bd:s:defense}

Attacks from white- to black-box settings have been developed to subvert the machine learning model to include backdoored behavior. However, any model trainer or holder may take proactive steps to detect and defend their models against this threat. This section will describe at length how this attack may be thwarted.
Overall, the task of detecting and defending against a backdoor attack can be divided into three key sub-tasks:

\begin{enumerate}
    \item {\bf Task 1:} \textit{Detecting the existence of the backdoor}. For a given
          model, it is difficult to know if the model is compromised
          (i.e., a model with a backdoor) or not. The first step of detecting
          and defending against the backdoor attack is to analyze the model
          and determine if there is a backdoor present in this model.
    \item {\bf Task 2:} \textit{Identifying the backdoor trigger}. When a backdoor is
          detected in a model, the second step is usually to identify which
          pattern (including its size, location, texture, and so on) is used as
          the trigger.
    \item {\bf Task 3:} \textit{Mitigating the backdoor attack}. After identifying the
          existence of a backdoor, the mitigation of such an attack is to
          remove the backdoor behavior from the model. Note that backdoor
          models can be made to be robust against transfer learning or fine-tuning~\cite{yao2019latent}.
\end{enumerate}

Note that not all detection and defense techniques will support all three 
sub-tasks. As some may assume prior knowledge that a
backdoor has already been detected, and the proposal only contains techniques to recover the trigger or mitigate the attack.

\begin{figure}
    \centering
    \includegraphics[width=\textwidth]{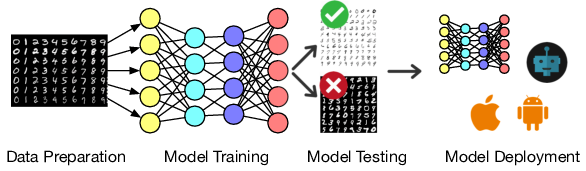}
    \caption{Overview of DNN model training and deployment.}
    \label{fig:overview}
\end{figure}

Figure~\ref{fig:overview} shows an overview of the DNN model training and
deployment process. It can be broken up into four general steps from data
preparation, model training, model testing, and model deployment. As discussed
in Section~\ref{c:bd:s:attacks}, most existing poisoning attacks target the
model training (or model retraining) step. Thus, investigating if the model contains a backdoor, reconstructing potential triggers, and/or mitigating any backdoor attacks must occur after this training step.
Thus, mitigation strategies will be employed either during model testing (i.e., pre-deployment) or
at the model's runtime (i.e., post-deployment), and hence, depending on when the inspection occurs, existing detection and defense techniques can be divided into two categories: pre-deployment techniques or post-deployment techniques.

\subsection{Pre-deployment Techniques}
\label{c:bd:s:defense:pre}

\subsubsection{No Inspections}
\label{c:bd:s:defense:pre:noinspection}

There exists work~\cite{zhao2018resilience, liu2018fine} attempting to directly
mitigate the backdoor attack without inspecting the model behavior. The key
technique behind these methods is to compress the model (e.g., by model
pruning or similar techniques) or fine-tune the model with benign inputs to
alter the model behavior hoping that the backdoor behavior is eliminated.
Specifically, Zhao et al.~\cite{zhao2018resilience} found that model pruning
can remove some behaviors of a trained model, and potentially it can remove
the backdoor of the model if pruning is purely using benign data.

Liu et al.~\cite{liu2018fine} observe that pruning the model alone does not
guarantee the removal of the model backdoor behavior. This is because the
malicious model may use the same neuron to demonstrate both benign and malicious
behaviors. 
Thus, if the neuron is removed, the model accuracy will be lower than that of
the original model. This would be an undesirable consequence even though the model backdoor is
removed. However, if this neuron is not pruned, the backdoor behavior is retained and the model continues to be malicious, also undesirable. 
Similarly, fine-tuning the model does
not necessarily remove the model backdoor, as some
attacks~\cite{yao2019latent} target transfer learning scenarios where
fine-tuning is needed. To solve this problem, Yao et al. propose Fine-Pruning, which combines the strengths of both fine-tuning and pruning to effectively nullify
backdoors in DNN models. Fine-Pruning first removes backdoor neurons using
pruning and fine-tuning the model in order to restore the drop in classification
accuracy on clean inputs (which is introduced in the previous pruning
procedure).

There are some limitations to these types of defenses. Firstly, model pruning
itself has unknown effects on the model. Even though model accuracy after
pruning does not decrease too much, many other important model properties, such
as model bias (sometimes known as fairness) and model prediction performance, are
not guaranteed to be the same. Using such models may potentially lead to
severe consequences. Secondly, these mitigation techniques assume access
to the training process and clean inputs, which conflicts with poisoning-based attacks.

\subsubsection{Pre-deployment Model Inspections}
\label{c:bd:s:defense:pre:static}

Before the model is deployed, it is possible to check whether the model has
been backdoored directly. This kind of strategy works without the running of
the model, so it is also called static detection. For these types of
techniques, some will require a large set of benign inputs to identify
backdoors, such as Neural Cleanse (NC)~\cite{wang2019neural}, whereas others
do not require much data (i.e., a limited number or even zero samples),
such as ABS~\cite{liu2019abs}.

\begin{figure}[ht]
    \centering
    \includegraphics[width=\textwidth]{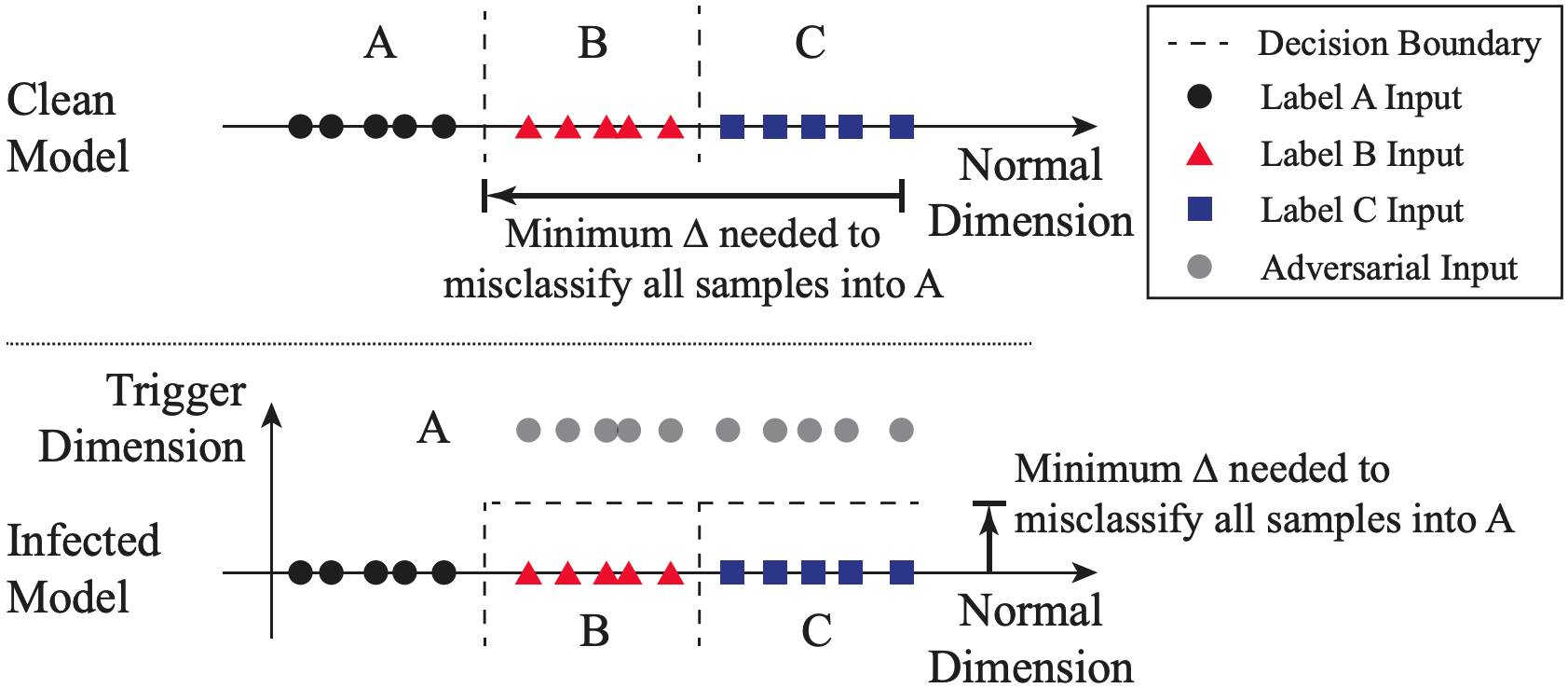}
    \caption{Intuition of NC~\cite{wang2019neural}.}
    \label{fig:nc}
\end{figure}

\noindent
{\bf Neural Cleanse.} Wang et al.~\cite{wang2019neural} propose Neural
Cleanse (NC), a pre-deployment technique to inspect DNNs, identify backdoors,
and mitigate such attacks. Figure~\ref{fig:nc} illustrates the key observation that enables NC. 
The top figure shows a clean model with three output labels. If we want to
perturb inputs belonging to C to A, more modification is needed to move
samples across decision boundaries. The bottom figure shows the infected model,
where the backdoor changes decision boundaries leading to a small perturbation value for changing inputs belonging to B and C to A.

Based on this observation, NC proposes to first compute a universal
perturbation, which is the minimized amount of change to make the model
predict a given target label. If the perturbation is small enough (i.e.,
smaller than a given threshold), NC considers it as one trigger. It then
verifies this by adding this trigger to a large number of benign inputs 
and tests if it is really a trigger and tries to optimize it based on prediction results.
After identifying the trigger, it can mitigate the attack by either using a
filter (i.e., to detect images with such a trigger pattern) or patching the
DNN by removing the corresponding behaviors by pruning the neural network.

NC has a number of limitations. First, NC makes an incorrect assumption that if
pixels in a small region have a strong influence on the output result, they
are treated as backdoor triggers. This results in NC confusing triggers with strong benign features.
In many tasks, there exist strong local features, where a region of pixels is
important for one output label, for example, the antlers of deers in CIFAR-10.
Secondly, NC assumes that the trigger has to be small and in the corner areas.
These are heuristics, which do not hold for many attacks. For example, Salem
et al.~\cite{salem2020dynamic} propose a dynamic attack, where triggers can be
added to different places and can successfully bypass NC. Thirdly, NC
requires a significant number of testing samples to determine if
a backdoor exists in a model or not. In real-world scenarios, such a large
number of benign inputs may not exist. Lastly, it is designed purely for input
space attacks, and it does not work for feature space attacks, such as using
Nashville and Gotham filters as triggers~\cite{liu2019abs}.

\begin{figure}
    \centering
    \includegraphics[width=0.31\textwidth]{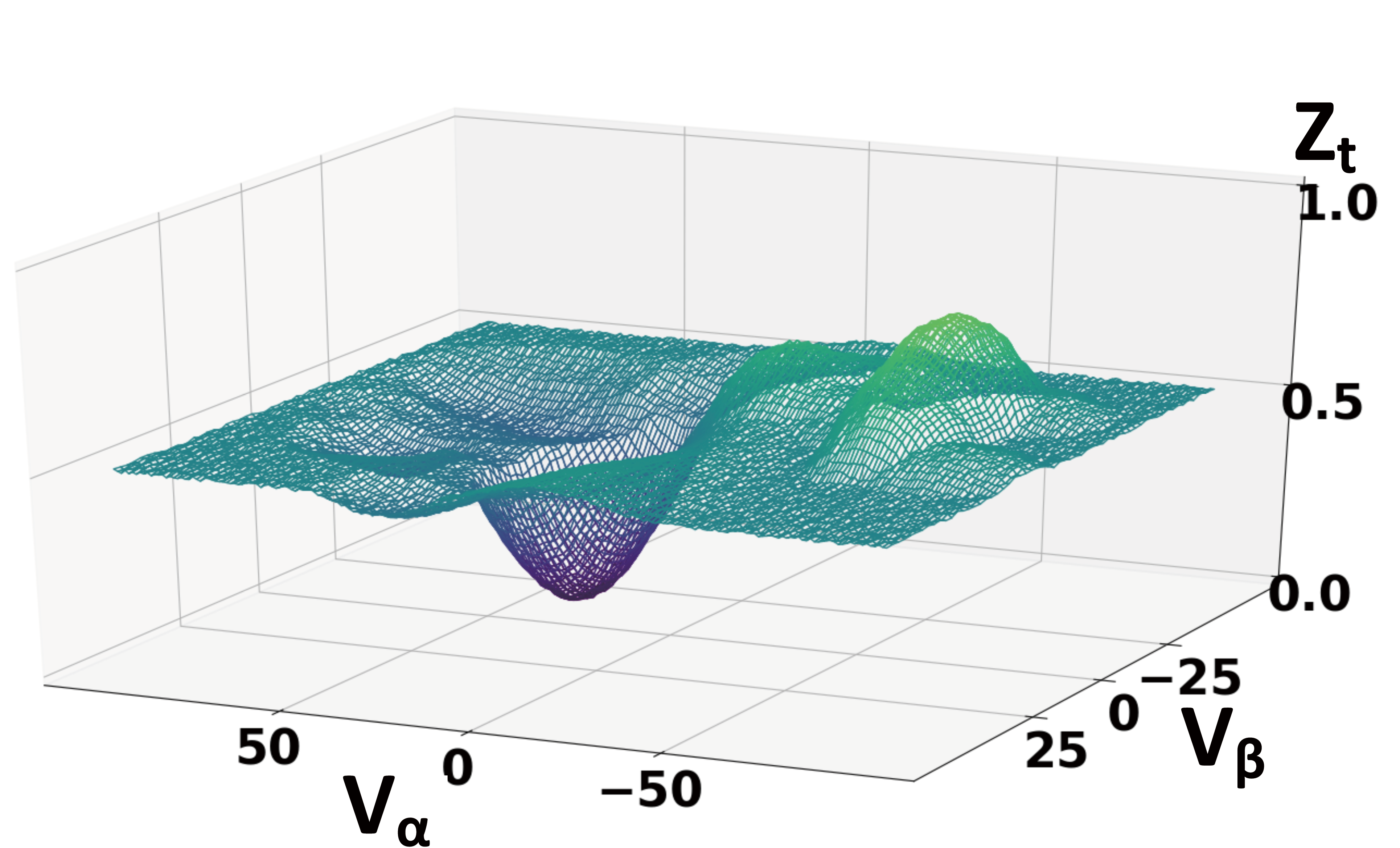}\quad
    \includegraphics[width=0.31\textwidth]{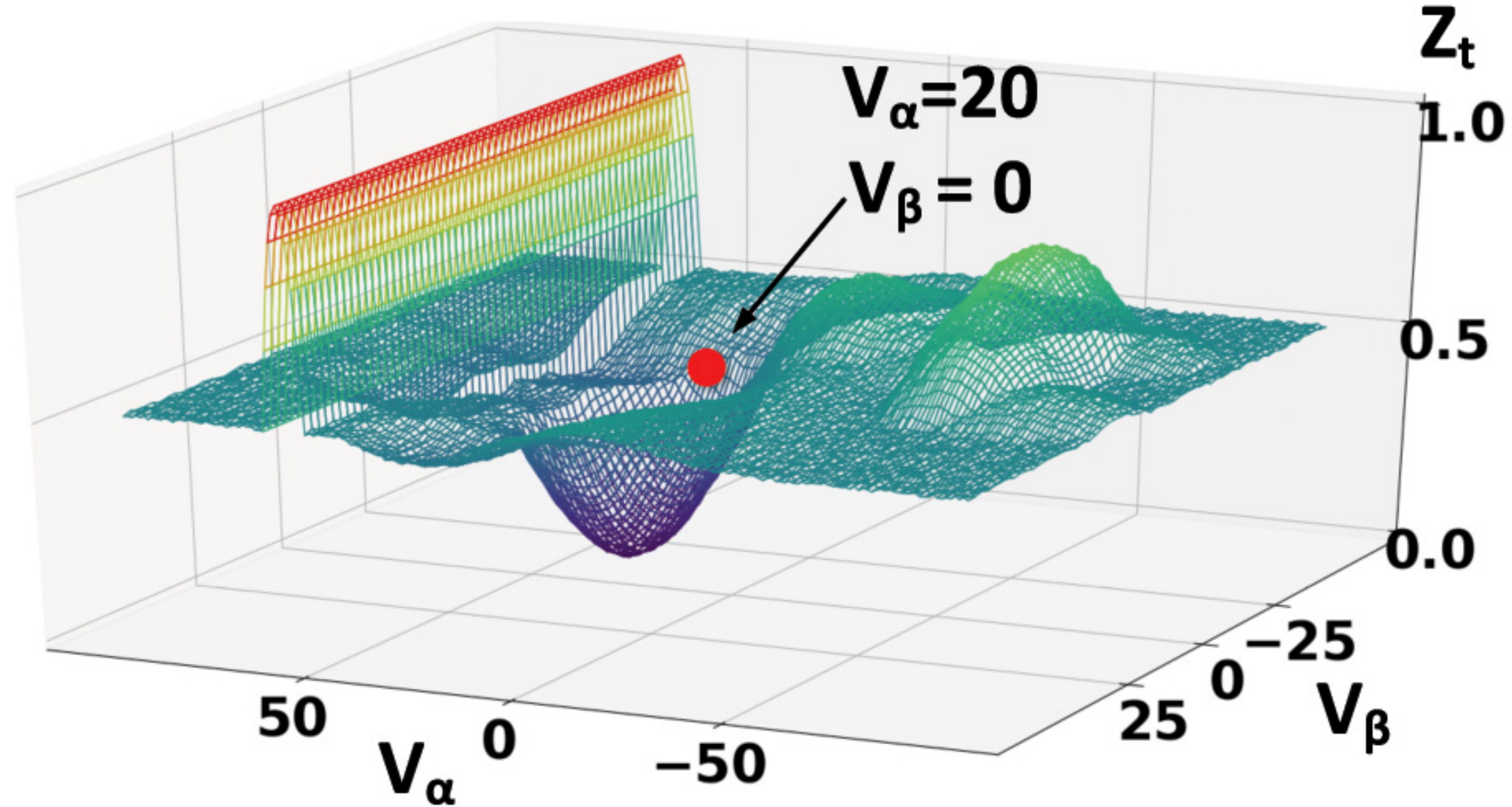}\quad
    \includegraphics[width=0.32\textwidth]{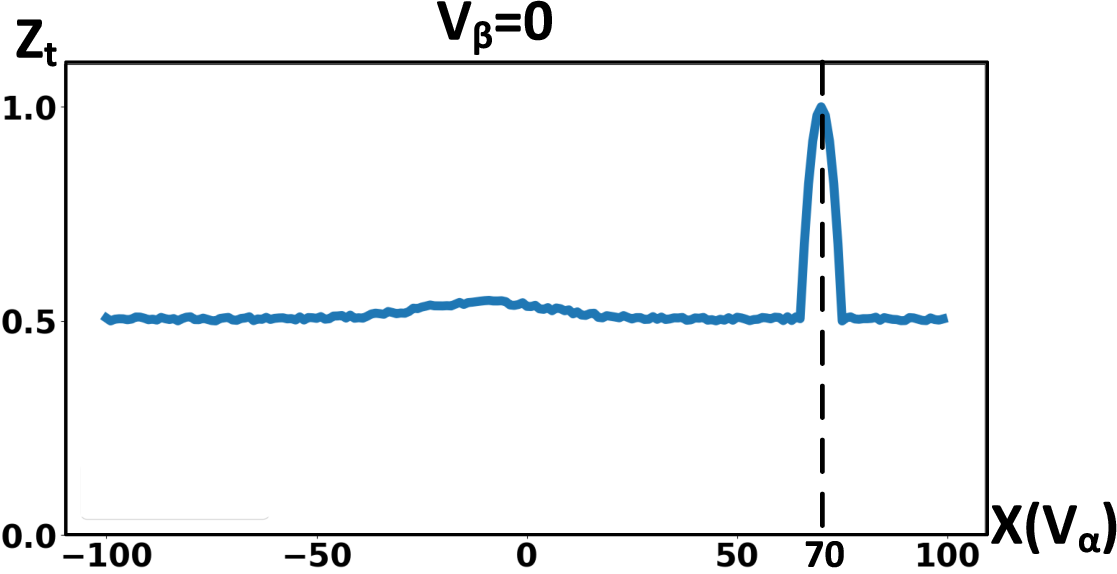}
        \vspace{0.3cm}
    \caption{Overview of ABS observations~\cite{liu2019abs}. The left figure
        shows the feature surface of a benign model. The middle figure shows the
        feature surface of a model with a backdoor. The right figure shows a slice
        of surface for the backdoored model. The red dot in the middle figure
        shows a state where the attack happens, and it corresponds to the dashed
        line in the right figure.}
    \label{fig:abs}
\end{figure}

\noindent
{\bf ABS.} ABS is built on top of two key observations. The first is that
successful attacks entail compromised neurons. In existing attacks, the
backdoored model recognizes the trigger as a strong feature of the target label to achieve a high attack success rate. Such a feature is represented
by a set of inner neurons, which are referred to as {\em compromised neurons}.
The second observation is that compromised neurons represent a subspace for
the target label that cutcrosses the whole space. This idea is shown in
Figure~\ref{fig:abs}. The feature space surfaces of a benign model (left figure in
Figure~\ref{fig:abs}) and that of a backdoored model are noticeably different. 
For a backdoored model, there exists a cut of the surface that is significantly
different from the benign model due to the injected backdoor. As it works
for all inputs, it will affect every prediction results once it is
activated. Thus, it will interact with the whole interface. The phenomenon is
demonstrated in the right figure of Figure~\ref{fig:abs}. When a neuron value
is assigned to a special value, i.e., the trigger pixel value, the output will significantly deviate from normal.

Based on these observations, Liu et al.~\cite{liu2019abs} propose Artificial
Brain Stimulation (ABS). For any given input, ABS first predicts its label
using the neural network. Then, it enumerates all neurons and performs a brain
stimulation process. Namely, for each neuron, it tries to change its
activation value to all possible values and simultaneously observes the
value changes in the output. If there is one neuron whose behavior is similar
to the right figure in Figure~\ref{fig:abs}, ABS treats it as a backdoor. To
reconstruct the backdoor trigger, ABS then performs a reverse engineering
process, which will try to find an input pattern that can strongly activate these compromised neurons and trigger the attack.

ABS also introduces a new type of backdoor attack, which is the feature space
attack. Namely, the trigger is no longer an input pattern (i.e., a region with
specific pixel values), but feature space patterns represent high-level
features (e.g., an image filter). However, this attack also has its own limitations.
Firstly, it assumes one backdoor for each class. This may not hold in
practice, and backdoors have been shown to be dynamic~\cite{salem2020dynamic}. Secondly, it currently enumerates neurons one by one, assuming the presence of a strong correlation between one neuron and the backdoor behavior, which may be hidden or overridden by more advanced attacks.

\subsection{Post-deployment Techniques}
\label{c:bd:s:defense:post}

In addition to static approaches functioning before models are deployed, there is
also work that monitors the model at runtime and determines if the
model has a backdoor and more importantly, if it has been triggered by an
input or not. In this setting, the defense or detection system can inspect individual inputs, offering a focused means of directly reconstructing 
the trigger by inspecting the attack input.

\begin{figure}
    \begin{center}
        \includegraphics[width=\textwidth]{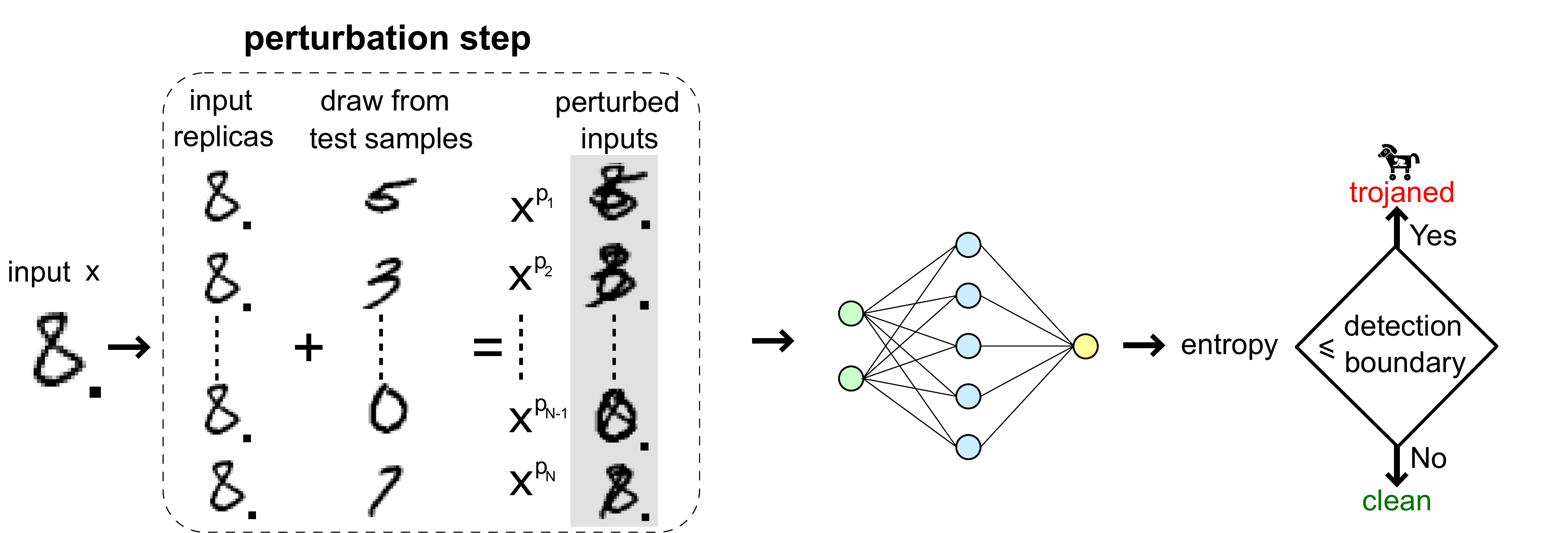}
    \end{center}
    \caption{Overview of STRIP~\cite{gao2019strip}.}
    \label{fig:strip}
\end{figure}

\noindent
{\bf STRIP.} Gao et al.~\cite{gao2019strip} propose STRong Intentional
Perturbation (STRIP), a run-time trojan attack detection system. The
workflow of STRIP is shown in Figure~\ref{fig:strip}. Firstly, STRIP will
perturb each input by adding benign samples drawn from the test samples to obtain
a list of perturbed inputs $X^{P_1}, X^{P_2}, . . ., X^{P_N}$. These inputs are the
overlap of a benign input and the given input. Next, it will feed all these
inputs to the DNN model. Note that if the input contains a trigger, it is highly
likely that a majority of the perturbed inputs will also yield predictions with 
the malicious output label result (due to the existence of the trigger), 
whereas for a benign input, the results are closer to random. 
As a result, STRIP only needs to examine every prediction result, and can then 
make a judgment on if the input will trigger the backdoor or not.

STRIP can effectively detect backdoor models and inputs that trigger the
backdoor if the trigger lies in the corners of the image or at least does not
overly overlap with the main contents. Such an example is shown in
Figure~\ref{fig:strip}). However, if the trigger does overlap with the
contents (e.g., overlap with digits in Figure~\ref{fig:strip}), the
detection will fail because the texture of the trigger will also be changed by the
perturbations. Salem et al.'s~\cite{salem2020dynamic} proposed a dynamic
backdoor attack that uses triggers that can be in the middle of the image.

\section{Applications of Backdoors}
\label{c:bd:s:apps}

\subsection{Watermarking} 
Digital Watermarking conceals information in a piece of media (e.g., sound, video, or images) to enable a party to verify the authenticity or the originality of the media. This watermark, however, must also be resilient to tampering and other actors seeking to subvert the legitimate piece of media.

Adi et al.~\cite{Adi2018Turning} propose an IP protection method for DNNs by applying the backdoor to watermark DNNs. They present cryptographic modeling for both tasks of watermarking and backdooring DNNs, and show that the former can be constructed from the latter (through a cryptographic primitive known as \textit{commitment}) in a black-box manner.
The definition of the backdoor attack that Adi et al. provided in a cryptographic framework is as follows: Given a trigger set $T$ and a labeling function $T_L$, the backdoor shall be termed as $b=(T, T_L)$. The backdooring algorithm $Backdoor(O^f, b, M)$ is a probabilistic polynomial-time (PPT) algorithm that receives as input an oracle to $f$ (ground-truth labeling function $f:D \rightarrow L$, where $D$ is input space, $L$ is output space), the backdoor $b$ and a model $M$, and outputs $\hat{M}$. $\hat{M}$ is considered $backdoored$ if
\begin{equation}
    \label{c:bd:s:apps:wtmk}
    \begin{aligned}
     Pr_{x \in \bar{D} \backslash T } [ f(x) \ne Classify(\hat{M}, x) ] \leq \epsilon ,\\
     Pr_{x \in T} [ T_L(x) \ne Classify(\hat{M}, x) ]  \leq \epsilon , \\
    \end{aligned}
\end{equation} 
where $\bar{D}$ is the meaningful input, $Classify(M,x)$ is a deterministic polynomial-time algorithm that, for an input $x \in D$ outputs a value $M(x) \in L \backslash \{ \perp \}$, and $\perp$ is an undefined output label. 
This definition presents two ways to embed a backdoor. The first is that the backdoor is implanted into a $\textit{pre-trained model}$. The second is the adversary can train a new model from scratch.

A watermarking scheme can be split into three key components. 
\begin{enumerate}
    \item Generation of the secret ``marking'' key $mk$. This key will be embedded as the watermark. A public verification key $vk$ is also generated and will be used later to detect the watermark. In watermarking via backdoors, the backdoor is the marking key, while a commitment (the cryptographic primitive) used to generate the backdoor is the verification key.
    \item Embedding the watermark (a backdoor $b$) into a DNN model. Through poisoned training data or retraining, as previously described in Section~\ref{c:bd:s:attack:tht1}, the watermark (backdoor) can be embedded.
    \item Verifying the presence of the watermark. Provided $mk, vk$, for a backdoor test $b=(T, T_L)$. If $\forall t^{(i)} \in T : T_L^{(i)} \ne f( t^{(i)} ) $, proceed to the next step, otherwise, the verification fails.
    Despite the detection of the watermark, one must verify the integrity of the commitment, i.e., if it was tampered or not. In the final step, the accuracy of the algorithm is verified. For all $i\in {1,..,n}$, if more than $\epsilon |T|$ elements from $T$ does not satisfy $Classify( t^{ (i) }, M ) = T_L^{(I)}$, then the verification fails, otherwise the commitment has been successfully verified.
\end{enumerate}

Adi et al.~\cite{Adi2018Turning} prove their method upholds the properties of:

\begin{itemize}
    \item \textit{Functionality-preserving}: the prediction accuracy of the model should not be negatively influenced by the presence of the watermark.
    \item \textit{Unremovability}: an adversary with full knowledge of the watermark generation process should not be able to remove the watermark from the model.
    \item \textit{Unforgeability}: an adversary with only the verification key should not be able to demonstrate ownership of the marking key.
    \item \textit{non-trivial ownership}: with knowledge of the watermark generation algorithm, a third party should not be able to generate marker and verification key pairs, and claim models for future models.
\end{itemize}

Li et al.~\cite{li2019persistent}, however, observe that the watermarking system proposed by Adi et al.~\cite{Adi2018Turning} makes the assumption that only one backdoor (watermark) may be inserted into the model. For example, Salem et al.~\cite{salem2020dynamic}'s Dynamic Backdoors contain multiple backdoors. The existence of multiple backdoors would result in multiple valid watermarks, and thus void the \textit{Unforgeability} claim. The insertion of multiple backdoors would also impact the \textit{Unremovability} of the original backdoor, otherwise termed as the persistence of the watermark.
In response, Li et al. leverage two data preprocessing techniques that use out-of-bound values and null-embedding to improve the persistence of the watermark against other attackers and limit the effects of retraining in the event that another backdoor is to be injected on top of the existing backdoor. 
Li et al. also introduce \textit{wonder filters}, a  primitive to enable the embedding of bit-sequences (from the marker key) into the model.

The largest hurdle to overcome in the application of the backdoor attack as a means to watermark DNNs, is that neural networks are fundamentally designed to be tuned and trained incrementally. Li et al. propose a \textit{model piracy} attack setting whereby an adversary wants to stake its own ownership claims on the model, or destroy the original owner's claims. To defend against this attack, Li et al. design a DNN watermarking system based on wonder filters that strongly authenticates owners by embedding (into the DNN) a filter described by the owner's private key.
Where Li et al's work differs from Adi et al. is in the \textit{wonder filter} $W$, which is a two-dimensional digital filter that can be applied to any input image. This filter will have 3 possible permutations for each pixel, transparent, positive change, or negative change, with a majority of filter pixels being transparent. Thus, $W$ is defined by the position, size, and values of a $0/1$ bit pattern block. 

When Li et al. apply \textit{out-of-bound values}, they translate the $0/1$ bit pattern of $W$ as out-of-bound values in the input images. A set of training data is processed with the filter. They then flip the values of the \textit{wonder filter} to create an \textit{inverted wonder filter} $W^-$. 
The \textit{inverted filter} $W^-$ is then applied to the same set of training data processed by $W$ 
The set of images filtered by $W$ are labeled as the target class label, while the $W^-$ filtered data is labeled as the original class label before the data is used to train the backdoored model. As for the normal and null embeddings approach, the normal and null embeddings serve complementary objectives. The normal embedding injects the desired marker into the model, while the null embedding ``locks down'' the model, so no additional watermarks may be added. 

Li et al's process of watermarking the image is similar to Adi er al's process, with the same three key processes of generating the secret ``marking'', or in this instance, the wonder filter $W$, embedding the watermark (and/or additionally locking down the model), and finally, the process of verifying the watermark, by using the image to compute $W$ and an associated label. After applying $W$ to a random set of images, it is expected that an authentic watermark should yield a majority of the target class label, instead of a random assortment of classes as expected from a random set of images, without $W$.

Li et al. also provide a security analysis to prove that their approach can uphold the requirements of \textit{reliability, no false positives, unforgeability,} and \textit{persistence}, whereby \textit{Reliability} describes that for a given input $x$, a poisoned input ($x \bigoplus W$, or $x \bigoplus W^-$), the backdoored DNNs will produce the predefined output in a deterministic manner. \textit{No False Positives} denotes that a verifier should not be capable of judging a clean model as the watermarked model. \textit{unforgeability} ensures that the watermark injected on a DNN has a strong association with its owner, and \textit{Persistence} guarantees that the watermark embedded cannot be corrupted or removed by an adversary.

\subsection{Adversarial Example Detection}

In Gotta Catch~\cite{shan2019gotta}, Shan et al. observe that the backdoor attack will alter the decision boundary of the DNN models. Following the injection of a backdoor, the decision boundary of the original clean model will mutate. The mutation will result in triggers establishing shortcuts in the decision boundary of the backdoored model. 

On the contrary, there is a common approach of adversarial attacks to find adversarial examples; for example, universal adversarial attacks~\cite{universal_AE, shafahi2018universal}, will try to iteratively search the whole dataset to find similar shortcuts to use as their universal adversarial examples. Based on this observation, the shortcut created by a backdoor can act as a trapdoor to capture the adversarial attacker's optimization process, detect, and/or recover from the adversarial attack~\cite{shan2020gotta}. The trapdoor implementation uses techniques similar to those found in BadNets backdoor attacks~\cite{gu2017badnets}. The authors define the trapdoor perturbation (the trigger) from multiple dimensions, e.g., mask ratio, size, pixel intensities, and relative locations.

\subsection{Open Problems}
\begin{itemize}
    \item \textit{Fair comparison of methods including attacks and defenses}. We have observed experimental settings (e.g. models and datasets) vary greatly across studies of different domains. Consequently to inspect the potential of AI Trojans in a standardized manner, The Intelligence Advanced Research Projects Activity (IARPA) has recently launched a program (and competition), Trojans in Artificial Intelligence (TrojAI)~\cite{trojAI} , .
    \item \textit{Persisting through fine-tuning in transfer learning.} A model backdoor can be made ineffective after only a few fine-tuning layers~\cite{gu2017badnets, yao2019latent}. Unfortunately, this limits the practicality of the backdoor attack in the transfer learning setting. More advanced backdoor attacks that can persist through such fine-tuning processes remains a challenge.
    \item \textit{Combining hidden triggers and clean labels.} To inject trojans into DNNs by poisoning the training data, the most stealthy method is to attach the triggers on the poisoned data in an imperceptible way, in addition to a correctly annotated label. Existing attacks either visually hide the triggers but retain an clearly poisoned label or correctly annotate the label with noticeable triggers. The combination of these stealth tactics, to produce poisoned data with an invisible trigger and clean labels still remains a challenge. 
    \item \textit{Accessing training data.} There are a variety of backdoor attacks, however only a small set has access to clean samples. In most security-sensitive cases, the attacker can only access a pre-trained model. The injection of a trojan by directly compromising the model weights without requiring access to the clean original training data still remains challenge.  
    \item \textit{Adaptiveness of Attacks and Defenses.} Existing attacks and defenses often discuss their effectiveness against reactive (dynamic) attack or defense countermeasures in a superficial and heuristic way, without including the adversary's possible countermeasures as a part of their works. These adaptive attacks and defenses can adaptively take optimal strategies when responding to an adversary's countermeasures, will provide improved practicality in realistic settings, but still remains an open challenge.
 
\end{itemize}
% \include{Contents/appendix}

% \backmatter%%%%%%%%%%%%%%%%%%%%%%%%%%%%%%%%%%%%%%%%%%%%%%%%%%%%%%%
% \include{Contents/glossary}
% \include{Contents/solutions}
% \printindex

\bibliographystyle{acm}
\bibliography{p.bib}
%%%%%%%%%%%%%%%%%%%%%%%%%%%%%%%%%%%%%%%%%%%%%%%%%%%%%%%%%%%%%%%%%%%%%%

\end{document}